# Symmetry-governed nonlinear metasurface simultaneously enhances circular dichroism, helicity purity, and Q-factor

Dmitrii Gromyko[1,2,#], Wenhe Jia[2,#], Chengfeng Pan[2,3,#], Xuan Mao[4,#], Wenduo Chen[2,4], Zhen Xu[4], Shutao Zhang[2,3], Hongtao Wang[5,6], Bochang Wu[2,3], Wang Hao[7,8], Sergei Tikhodeev[9], Qijie Wang[4,10,11,*], Joel K.W. Yang[3,*], Cheng-Wei Qiu[2,*], Lin Wu[1,*]

[1]Science, Mathematics, and Technology, Singapore University of Technology and Design, 8 Somapah Road, 487372, Singapore

[2]Department of Electrical and Computer Engineering, National University of Singapore, 4 Engineering Drive 3, 117583, Singapore

[3]Engineering Product Development, Singapore University of Technology and Design, 8 Somapah Road, 487372, Singapore

[4]School of Electrical and Electronic Engineering, Nanyang Technological University, 639798, Singapore

[5]Beijing Engineering Research Center of Mixed Reality and Advanced Display, School of Optics and Photonics, Beijing Institute of Technology, Beijing 100081, China

[6]State Key Laboratory of Chips and Systems for Advanced Light Field Display, School of Physics, Beijing Institute of Technology, Beijing 100081, China

[7]Hangzhou International Innovation Institute, Beihang University, Hangzhou 311115, China

[8]School of Instrumentation and Optoelectronic Engineering, Beihang University, Beijing 100191, China

[9]Faculty of Physics, Lomonosov Moscow State University, Leninskie Gory 1, Moscow 119991, Russia

[10]Centre for Disruptive Photonic Technologies, Nanyang Technological University, 637371, Singapore

[11]School of Physical and Mathematical Sciences, Nanyang Technological University, 637371, Singapore

[#]These authors contributed equally to this work

*Correspondence: qjwang@ntu.edu.sg, joel_yang@sutd.edu.sg, chengwei.qiu@nus.edu.sg, lin_wu@sutd.edu.sg

# Abstract

Circularly polarized light underpins a broad range of chiroptical and spin-selective photonic technologies, yet extending resonant chirality to nonlinear harmonic emission remains challenging. Here, we experimentally demonstrate a symmetry-governed mechanism that maps the maximal linear chirality of high-Q degenerate resonances onto the polarization of the generated harmonic. Preserved rotational symmetry protects modal degeneracy and polarization-selection rules, while broken mirror symmetries enable a strong chiroptical response. The metasurface thereby filters the pump polarization and purifies the handedness of the nonlinear output, simultaneously achieving third-harmonic circular dichroism of −0.98, a degree of circular polarization of 0.83, and an effective nonlinear Q-factor of 140. Importantly, the nonlinear emission preserves its handedness even under spatially depolarized excitation, demonstrating robustness to variations in the incident polarization state. These results establish complementary symmetry preservation and breaking as a deterministic route for transferring resonant chirality from linear to nonlinear optical responses, providing a general design principle for nonlinear chiral photonics.

Circularly polarized (CP) light plays a central role in modern photonics, being a key element of chiroptical spectroscopy[1,2], spin–orbit photonics[3,4], spintronics[5], and quantum optics[6]. The growing demand for compact, on-chip sources of CP radiation has sparked a rapid development of chiral metasurfaces: nanostructured optical devices capable of tailoring light's handedness[7,8], emission direction[9-11], and angular momentum[12,13]. Chiral metasurfaces have been used in a variety of applications, such as the creation of single-handed optical cavities[14], spin multiplexing of single photons[15], chiral lasing[16-20], and spin-direction-locked emission[9,21].

While these advances exploit linear light-matter interactions to control polarization and emission, extending such control into the nonlinear regime remains a major challenge. In this context, dielectric nonlinear metasurfaces[22-25] have emerged as ultra-thin photonic platforms for frequency conversion, where strong resonant field enhancement due to high-Q resonant modes enables efficient generation of the second[26-28], third[29-31], and higher harmonics[32,33] within patterned slabs of subwavelength thickness. Despite the recent progress in conversion efficiency[28], mastering the chirality of nonlinear radiation remains an outstanding challenge. Importantly, while early nonlinear metasurfaces made of plasmonic materials demonstrated efficient polarization control[34-36], they suffered from strong ohmic losses, were limited in the conversion efficiency, and relied on surface-induced nonlinearity. Unlike in plasmonic metasurfaces, harmonic generation in dielectric metasurfaces arises from near fields with complex spatial structure within the volume of the nonlinear material. Such volume nonlinear response depends sensitively on geometry, material properties, and fabrication imperfections. Accordingly, most nonlinear chiral devices developed to date operate based on two main physical mechanisms.

The first mechanism is the well-known symmetry selection rule, stating that metasurfaces exhibiting a proper rotational symmetry generate CP nonlinear signals under normally incident right- and left-circularly polarized (RCP and LCP) excitation[37-41]. Despite the ability to produce CP nonlinear outputs, such structures are inherently achiral, and the chirality of the nonlinear signal is strictly conditioned by the excitation chirality. The second mechanism, which forms the basis for devices known as nonlinear chiral metasurfaces, exploits a chiral resonance at the excitation frequency[42-45]. Such resonances selectively couple to a specific CP excitation, producing near-field enhancement within the nonlinear material and generating a strong nonlinear signal. Excitation with the opposite CP, in contrast, produces only a weak, non-resonant output. Importantly, this handedness-dependent intensity asymmetry, commonly quantified by nonlinear circular dichroism, reflects only the difference in signal strength and does not convey information about the polarization state of the generated harmonic. At the same time, reduced symmetry of chiral metasurfaces does not provide any reliable mechanism for the polarization control of the nonlinear signal. The induced nonlinear near-fields, characterised by rapid spatial variations of magnitude, direction, and phase, result in the harmonic emission with arbitrary elliptical polarization. Consequently, no existing mechanism combines chirality with high-Q resonances to deterministically generate nonlinear harmonics with helicity fixed over a broad family of input states.

Only recently have CP nonlinear signals from dielectric metasurfaces been observed under specific linearly polarized (LP) excitations[46-48], but these effects arise from accidental modal degeneracies at the harmonic frequency, leading to polarization states that depend sensitively on modal configuration and excitation polarization. As we proposed in Ref.[49], a **deterministic route** requires a symmetry-protected mechanism based on **maximally chiral**, **high-Q** **resonances** in rotationally symmetric metasurfaces, where degenerate modes selectively couple to a chosen circular polarization and generate CP harmonics robustly. However, such resonances have been difficult to realize experimentally due to reliance on perfect vertical alignment in the double-layer geometry. On the other hand, existing demonstrations of high-Q chiral resonances, particularly quasi-bound states in the continuum (quasi-BICs), typically rely on complete symmetry breaking[8,19,42,50], which lifts degeneracy and precludes symmetry-protected polarization manipulation.

In this work, we overcome these limitations and enable practical implementations by introducing and experimentally verifying a symmetry-based mechanism for the generation of a chiral nonlinear signal, which can be interpreted as nonlinear transfer of resonant chirality. Our approach relies on a dielectric metasurface that combines nonlinearity-matching rotational symmetry with tailored mirror asymmetry to produce a pair of maximally chiral, high-Q quasi-guided resonances, implemented in a flat, single-layer, CMOS-compatible silicon platform (Fig. 1a,b). The helicity-selective coupling of the degenerate resonance pair imprints a specific chiral phase pattern onto the fields induced at the pump frequency. This phase is transferred to the third harmonic (TH) polarization through the rotational-symmetry selection rules, resulting in CP nonlinear harmonic radiation. We demonstrate that devices built on this principle simultaneously achieve near-unity TH circular dichroism of −0.98, TH degree of circular polarization (DCP) of 0.83 and high Q-factor controlled by the symmetry-breaking strength, thereby occupying a distinct performance regime in Fig. 1c. Moreover, we demonstrate that TH response remains dominated by the selected CP even after the excitation beam is transformed by a liquid crystal depolarizer (Fig. 1d). Establishing the symmetry-protected transfer of resonant chirality as a general design principle, this work lays the foundation for a new class of functional nonlinear devices with robust structurally-controlled polarization of nonlinear signal.

## Nonlinear chirality transfer mechanism

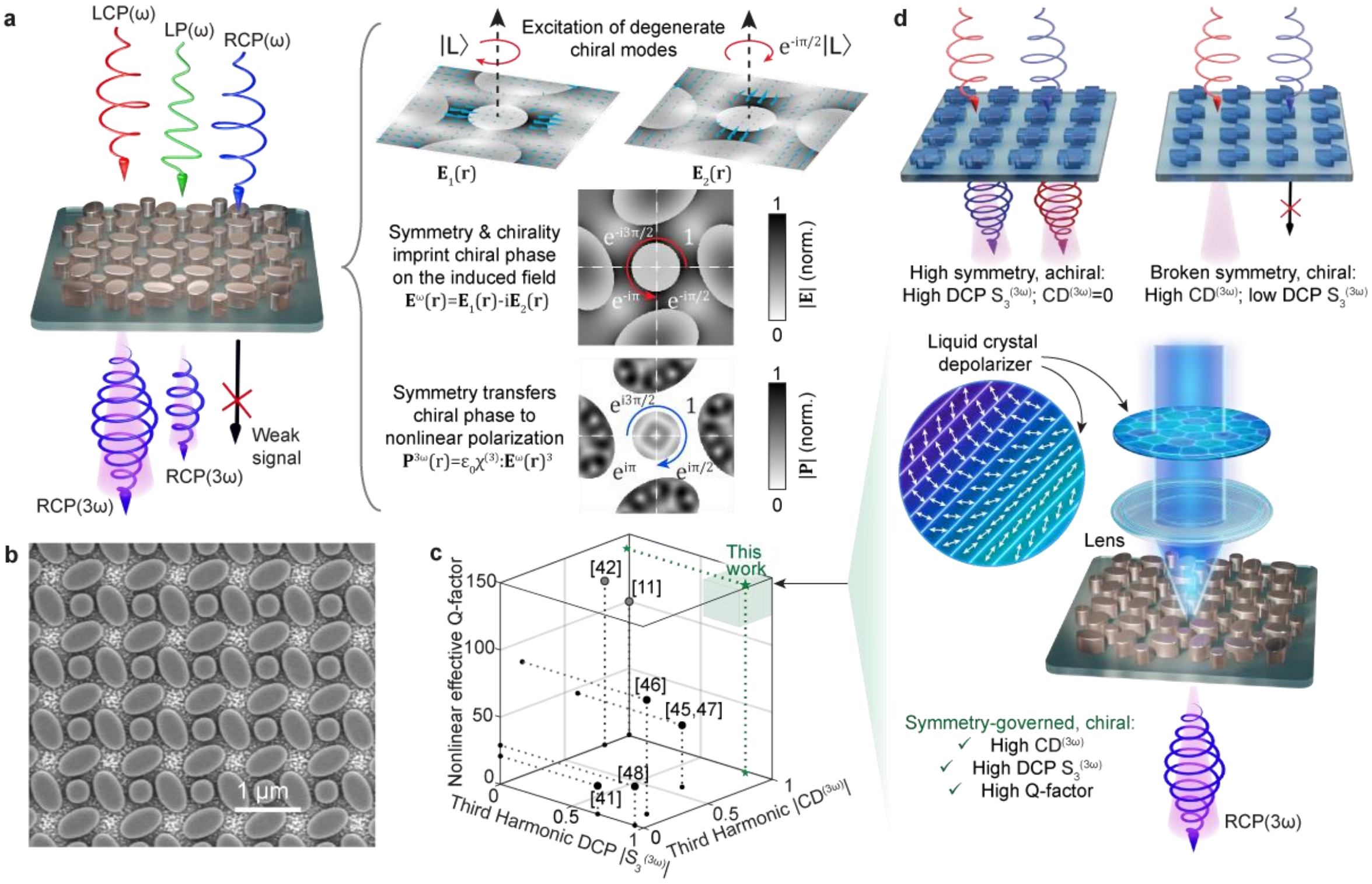

**Figure 1. Nonlinear chirality transfer in flat dielectric metasurfaces. a,** Symmetry-encoded chiral near-field pattern is transferred to the nonlinear polarization, enforcing robust generation of a circularly polarized TH signal. **b,** Scanning electron microscope image of the fabricated chiral metasurface. **c,** Quantitative comparison of experimental performance of all-dielectric nonlinear metasurfaces in terms of TH degree of circular polarization, nonlinear circular dichroism, and nonlinear Q-factor. Refs. [11,42] (gray circles) do not report TH polarization. **d,** Comparison of the established approaches for nonlinear chirality and generation of chiral TH signal under excitation with a spatially varying polarization state.

We consider a metasurface exhibiting fourfold rotational symmetry $C_4$ of the meta-lattice made of materials with nonlinear response that is invariant under $C_4$ (Fig. 1a). The symmetry is chosen to demonstrate the effect for TH generation; however, the phenomenon is general and can be realized in systems with $N$-fold rotational symmetry ($N = 3,4,6$) for the $(N-1)$th-order nonlinear process.

In structures with $N \geqslant 3$, it is possible to identify pairs of degenerate resonant modes, whose degeneracy is protected by the rotational symmetry of the structure. In the particular case, a mode pair with electric fields $\mathbf{E}_1(\mathbf{r}), \mathbf{E}_2(\mathbf{r})$ form the real two-dimensional irreducible representation $E$ of the $C_4$ point group, such that the two modes are converted into each other under a 90° rotation (Fig. 1a, right side):

$$\hat{C}_4 \mathbf{E}_1(\hat{C}_4^{-1}\mathbf{r}) = -\mathbf{E}_2(\mathbf{r}), \qquad \hat{C}_4 \mathbf{E}_2(\hat{C}_4^{-1}\mathbf{r}) = \mathbf{E}_1(\mathbf{r}). \tag{1}$$

As we show later, the far-field polarization of such modes can be controlled by means of geometric perturbations. Throughout this work, modal chirality refers specifically to the degree of circular polarization of the resonance's radiative far field (see Methods), quantified by $S_3$, rather than to local optical-chirality density. Once one of the modes is tuned to maximal chirality on the excitation side (LCP for definiteness), Eqs. (1) necessitate that the other mode exhibits the same polarization with a phase difference $\pi/2$. Thus, under arbitrarily polarized excitation, both modes are excited with equal amplitudes proportional to the LCP component of the incident field while preserving the polarization-independent phase offset (see Methods). Assuming that the non-resonant response remains weak, the induced electric field at the pump frequency, given by the combination of excited resonant modes $\mathbf{E}^{\omega}(\mathbf{r}) \propto \mathbf{E}_1(\mathbf{r}) - i\mathbf{E}_2(\mathbf{r})$, features a highly symmetric field distribution with an imprinted chiral phase pattern:

$$\hat{\boldsymbol{C}}_4 \mathbf{E}^{\omega}(\hat{\boldsymbol{C}}_4^{-1}\mathbf{r}) = \mathbf{E}^{\omega}(\mathbf{r})e^{-i\pi/2}. \quad (2)$$

This response is fundamentally different from that of achiral $C_{4v}$ structures, where degenerate modes with orthogonal far-field polarizations produce markedly different near-field patterns under CP and LP excitation (see Supplementary Fig. S1).

Resonant field $\mathbf{E}^{\omega}(\mathbf{r})$ induces nonlinear polarization $\mathbf{P}^{3\omega}(\mathbf{r}) = \varepsilon_0 \chi^{(3)}(\mathbf{r})\colon \mathbf{E}^{\omega}(\mathbf{r})^{\mathbf{3}}$, which, due to the invariance under $C_4$ rotations, follows similar transformation rules:

$$\hat{\boldsymbol{C}}_4 \mathbf{P}^{3\omega}(\hat{\boldsymbol{C}}_4^{-1}\mathbf{r}) = \mathbf{P}^{3\omega}(\mathbf{r})e^{-3i\pi/2}. \quad (3)$$

The phase factor $e^{-3i\pi/2}$ is equivalent to $e^{i\pi/2}$; therefore, the nonlinear polarization transforms as an RCP-induced field and consequently radiates an RCP TH signal into the lower half-space (see Supplementary Section S1 for details).

As a consequence, the chirality of the linear resonant response enforces the chiral transformation rules for the induced electromagnetic fields $\mathbf{E}(\mathbf{r})$ under an arbitrarily polarized pump. This chirality is subsequently transferred to the nonlinear polarization. The net result is a robust generation of an RCP nonlinear signal whose intensity is controlled by the pump polarization (Fig. 1a).

## Degenerate high-Q chiral modes in a flat metasurface

Demonstration of nonlinear chirality transfer requires a degenerate pair of modes of maximal chirality and sufficient resonant-to-nonresonant enhancement. Chiral quasi-BICs, which combine ultrahigh-Q resonances with fully CP emission, have attracted considerable interest since their theoretical introduction several years ago by two groups[7,50]. Subsequent experiments have demonstrated impressive Q-factors and near-perfect linear chiral response[8,19]. However, reported realizations rely on slanted[51], multilayer[10], multi-height[52,53], or even flat[42] architectures that break not only mirror but also rotational symmetries.

This reliance on asymmetric designs originates from fundamental symmetry selection rules[54,55]. In structures with four-fold rotational symmetry, these rules prohibit the formation of quasi-BIC resonances at normal incidence: all non-degenerate modes become symmetry-protected BICs, while degenerate modes necessarily couple to free space. Consequently, realizing a transition from a BIC to a quasi-BIC requires breaking rotational symmetry. In contrast, rotational symmetry is essential in our work to enable nonlinear chiral effects, rendering the quasi-BIC approach fundamentally incompatible with our requirements. Instead, we exploit rotationally symmetric, degenerate quasi-guided modes[56], *i.e.*, descendants of guided modes below the light line, which offer tunable high Q-factors[57] without sacrificing structural symmetry.

We introduce a general design principle for realizing degenerate high-Q chiral modes through sequential symmetry breaking and illustrate it with the example shown in Fig. 2. The initial structure, which supports guided modes, is shown in Fig. 2a. It consists of cylindrical amorphous silicon pillars with radius $R$ and height $H$, arranged in a square lattice on a $SiO_2$ substrate with period $p/\sqrt{2}$. While substrate-superstrate asymmetry has been shown to be essential for achieving a chiral response in rotationally symmetric structures, here we treat it as an inherent feature of a practically realizable metasurface[58]. The substrate is naturally retained to minimize fabrication complexity, and the design freedom is confined to the metasurface pattern. The corresponding modal dispersion is shown in Fig. 2b, with the first Brillouin zone and its high-symmetry points defined in Fig. 2c. Multiple modes appear near 200 THz at the $\Gamma^{(1,0)}$ point below both the vacuum and substrate light lines. Owing to their large Bloch wavevectors, these modes lie outside the radiation continuum and therefore possess infinite Q-factors.

To enable free-space coupling, we perturb the structure by introducing additional smaller cylindrical pillars with radius $R_c$, as shown in Fig. 2d, thereby forming a larger unit cell with period $p$. The expansion of the unit cell in real space results in Brillouin-zone folding in momentum space (Fig. 2c), which maps the guided modes from $\Gamma^{(1,0)}$ to the original Γ point at normal incidence. This process gives rise to a family of quasi-guided modes near 200 THz, as shown in Fig. 2e. Due to the symmetry-enforced degeneracy, such modes have been used to achieve pump-polarization-independent enhancement of the TH signal[59]; however, helicity control remained elusive. Similar to quasi-BICs, these modes can exhibit ultrahigh Q-factors that are tunable via the perturbation strength (Fig. 2f). Despite this similarity, quasi-guided modes are fundamentally distinct from quasi-BICs. In quasi-BIC systems, a radiative resonance persists in the dispersion even in the absence of perturbation and remains coupled to free space at oblique incidence. In contrast, the far-field coupling of quasi-guided modes vanishes throughout the entire momentum space as the perturbation strength approaches zero, as evident from Fig. 2b.

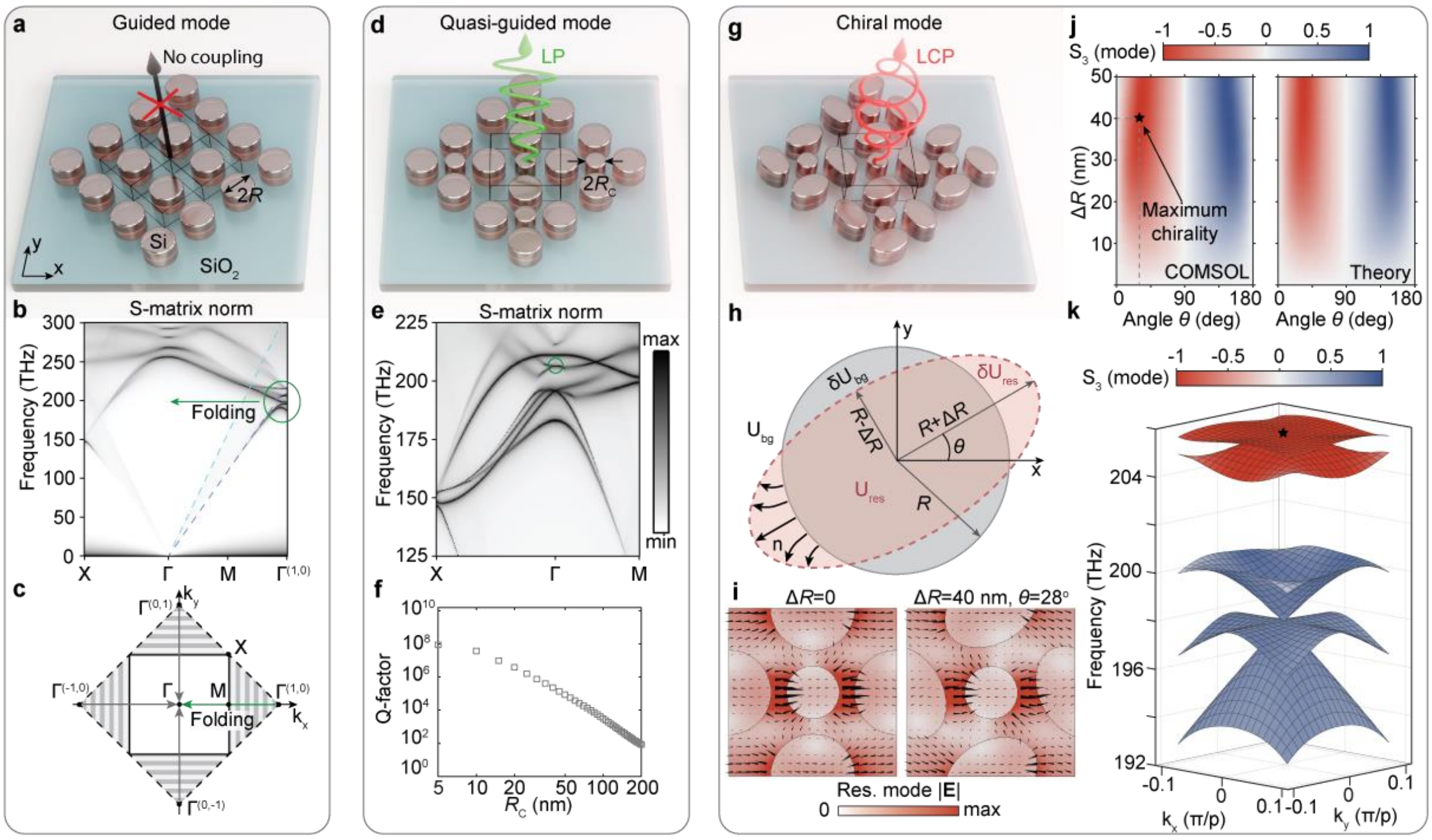

**Figure 2 | Enabling degenerate high-Q chiral modes via sequential symmetry breaking.** **a,** Initial structure supporting nonradiative guided modes below the substrate light cone near 200 THz. **b,** Scattering-matrix norm revealing the modal dispersion; dashed lines indicate the air and substrate light cones. **c,** Brillouin-zone folding of the guided modes into the first Brillouin zone. High-symmetry points are labelled according to the perturbed reciprocal lattice. **d,** Perturbed structure with an enlarged unit cell containing a smaller central cylinder of radius $R_c$. **e,** Dispersion of the resonant modes; the degenerate quasi-guided mode pair near 205 THz, arising from Brillouin-zone folding, is highlighted by the green circle. **f,** Q factor of the degenerate quasi-guided modes as a function of the perturbation parameter $R_c$. **g,** Metasurface with broken mirror symmetries hosting a pair of degenerate high-Q chiral modes. **h,** Schematic of the symmetry-breaking process: larger circular pillars are transformed into elliptical ones. **i,** Evolution of the resonant electric-field distribution during the transition to chirality. **j,** Mode chirality, quantified by the far-field degree of circular polarization, obtained from COMSOL simulations (left) and predicted by the interface perturbation theory (right). **k,** Dispersion of the resonant modes near the $\Gamma$ point at 200 THz. Colour denotes chirality; the black star marks two degenerate modes with maximum chirality.

We select a pair of degenerate quasi-guided modes near 205 THz, indicated by the green circle in Fig. 2e. These modes are initially achiral because the structure preserves in-plane mirror symmetries. To induce chirality, we morph the larger circular pillars into elliptical ones with half-lengths $R \pm \Delta R$, rotated by an angle $\theta$ (Fig. 2g), thereby breaking all mirror symmetries and introducing geometric chirality. The resulting chiral optical response, however, depends sensitively on the perturbation parameters $\Delta R$ and $\theta$. To analyze this dependence, we partition space into four regions defined by the permittivity distributions of the unperturbed structure $\varepsilon(\mathrm{r})$ and the perturbed structure $\tilde{\varepsilon}(\mathrm{r})$ in Fig. 2h. The regions $U_{\mathrm{res}}$ and $U_{\mathrm{bg}}$ retain the same permittivities, $\varepsilon_{\mathrm{Si}}$ and $\varepsilon_{\mathrm{air}}$, before and after perturbation. In contrast, the regions $\delta U_{\mathrm{res}}$ and $\delta U_{\mathrm{bg}}$ undergo permittivity exchange, switching from $\varepsilon_{\mathrm{air}}$ to $\varepsilon_{\mathrm{Si}}$ and from $\varepsilon_{\mathrm{Si}}$ to $\varepsilon_{\mathrm{air}}$, respectively.

To predict the values of $\Delta R$ and $\theta$ that yield maximum resonance chirality, we employ a semi-analytical model based on a perturbative approach and Maxwell's boundary conditions. At any material interface, the tangential components of the electric field and the normal components of the electric displacement field are continuous, *i.e.*, $\mathbf{E}_{1||} = \mathbf{E}_{2||}$ and $\mathbf{D}_{1\perp} = \mathbf{D}_{2\perp}$. As a result, the normal component of the electric field is discontinuous across Si–air boundaries. Small structural perturbations that shift these boundaries do not substantially modify the global field distribution, but they induce pronounced changes in the electric field within regions of altered permittivity, $\delta U_{\text{res}}$ and $\delta U_{\text{bg}}$, as can be seen in Fig. 2i. Figure 2i shows the evolution of just one mode in the degenerate pair, which in the achiral structure ($\Delta R = 0$ nm and $\theta = 0°$) exhibits pure $x$-polarization. The second mode, being a 90°-rotated copy of the first mode, undergoes an identical transformation.

While boundary-shift corrections have been shown to be essential for accurately capturing the energy shifts of perturbed resonant modes[60,61], we exploit them to evaluate the resonance far-field polarization and thus its chirality (see Methods for details). Essentially, the normal components of the resonant fields of the achiral structure are modified by the ratio of the permittivity distributions $\varepsilon(\mathrm{r})/\tilde{\varepsilon}(\mathrm{r})$ , such that the perturbed fields remain continuous wherever $\tilde{\varepsilon}(\mathrm{r})$ is continuous and satisfy Maxwell's boundary conditions at all interfaces of the perturbed structure. A comparison between the exact resonant mode chirality extracted from COMSOL eigenmode simulations (see Methods for details) for varying $\Delta R$, $\theta$ and the semi-analytical calculations based on the perturbed resonant fields indicates good agreement, as shown in Fig. 2j.

We choose a metasurface design with perturbation parameters $\Delta R = 40\,$nm and $\theta = 28°$, which correspond to the maximum chirality of the resonant mode. The resulting two-dimensional dispersion of modes near 200 THz is shown in Fig. 2k. The modes of interest are degenerate at the Γ point, are well separated from other bands, and exhibit near-flat dispersion. Importantly, they also display near-unity DCP over a substantial region of momentum space (indicated by colour in Fig. 2k), providing ideal conditions for observing nonlinear chirality transfer. This mechanism is first validated in numerical simulations and subsequently demonstrated experimentally in a silicon-on-silica platform.

## Numerical simulations of the nonlinear chirality transfer

Next, we analyse the linear and nonlinear optical responses of the metasurface tuned to exhibit maximum chirality with $\Delta R = 40$ nm and $\theta = 28°$ . Figure 3a,b compares the linear transmission spectra $T_{\mathrm{R,L}}$ under RCP and LCP excitation, together with the corresponding near-field enhancement spectra for both lossless and lossy structures. Remarkably, in the ideal lossless case, the resonant mode with maximal chirality is completely absent from the far-field scattering spectra. In contrast, the second resonant mode manifests as a sharp transmission dip but exhibits zero linear circular dichroism, defined as $\mathrm{CD} = (T_{\mathrm{R}} - T_{\mathrm{L}})/(T_{\mathrm{R}} + T_{\mathrm{L}})$ . This behaviour arises from the combined constraints of reciprocity and rotational symmetry, as established in prior theoretical studies[7,58].

Importantly, the absence of resonant features or linear CD in the scattering spectra does not imply suppression of the resonance or lack of polarization selectivity. On the contrary, the near-field intensity (Fig. 3a, bottom panel; Fig. 3c) reveals strong resonant enhancement with pronounced polarization dependence. When material losses are introduced to reach the critical-coupling regime (extinction coefficient $k = 0.01$) the same structure exhibits two prominent chiral resonances in transmission. While these losses maximize the depth of the resonant transmission dip, they have minimal impact on the resonance chirality and primarily reduce the near-field enhancement (Fig. 3b, bottom panel). This observation suggests that in rotationally symmetric structures, strong nonlinear circular dichroism can emerge even in the absence of strong linear CD, which we also observe experimentally. We note that this mechanism is different from the nonlinearity-induced chirality observed in linearly achiral resonant metasurfaces[62].

Moving on to nonlinear effects, Fig. 3d summarizes the two key measures of nonlinear chirality. First, we denote the RCP and LCP components of the generated TH signal under pump with polarization state $\boldsymbol{s}^{\text{in}}$ as $I^{(3\omega)}_{\boldsymbol{s}^{\text{in}}\rightarrow \text{R,L}}$. Nonlinear CD quantifies the dependence of the total TH intensity $I^{(3\omega)}_{\text{tot}} = I^{(3\omega)}_{\text{R}} + I^{(3\omega)}_{\text{L}}$ on the helicity of the incident field and is therefore obtained by comparing two separate measurements under RCP and LCP excitation:

$$\text{CD}^{(3\omega)} = \frac{I^{(3\omega)}_{\text{R}\rightarrow\text{tot}} - I^{(3\omega)}_{\text{L}\rightarrow\text{tot}}}{I^{(3\omega)}_{\text{R}\rightarrow\text{tot}} + I^{(3\omega)}_{\text{L}\rightarrow\text{tot}}} = \frac{I^{(3\omega)}_{\text{R}\rightarrow\text{R}} + I^{(3\omega)}_{\text{R}\rightarrow\text{L}} - I^{(3\omega)}_{\text{L}\rightarrow\text{R}} - I^{(3\omega)}_{\text{L}\rightarrow\text{L}}}{I^{(3\omega)}_{\text{R}\rightarrow\text{R}} + I^{(3\omega)}_{\text{R}\rightarrow\text{L}} + I^{(3\omega)}_{\text{L}\rightarrow\text{R}} + I^{(3\omega)}_{\text{L}\rightarrow\text{L}}}. \quad (4)$$

Without additional symmetry constraints, nonlinear CD merely describes how strongly the nonlinear conversion efficiency depends on the incident helicity but does not specify the polarization state of the generated TH signal. As one can see, near-unitary $\text{CD}^{(3\omega)}$ can be achieved even when the generated TH is linearly polarized, e.g. when $I^{(3\omega)}_{\text{R}\rightarrow\text{R}} = I^{(3\omega)}_{\text{R}\rightarrow\text{L}} \gg I^{(3\omega)}_{\text{L}\rightarrow\text{L,R}}$. We demonstrate that such behaviour is ubiquitous by considering the nonlinear response of some well-known chiral metasurfaces[11,42] in Supplementary Section S2.

On the contrary, the TH DCP, which quantifies the helicity purity, characterizes the polarization state of the generated nonlinear signal and is defined using the RCP and LCP TH components $I^{(3\omega)}_{\boldsymbol{s}^{\text{in}}\rightarrow \text{R,L}}$, produced in a single experiment under a fixed excitation state $\boldsymbol{s}^{\text{in}}$:

$$\text{DCP}^{(3\omega)} \equiv S^{(3\omega)}_3 = \frac{I^{(3\omega)}_{\boldsymbol{s}^{\text{in}}\rightarrow\text{R}} - I^{(3\omega)}_{\boldsymbol{s}^{\text{in}}\rightarrow\text{L}}}{I^{(3\omega)}_{\boldsymbol{s}^{\text{in}}\rightarrow\text{R}} + I^{(3\omega)}_{\boldsymbol{s}^{\text{in}}\rightarrow\text{L}}}. \quad (5)$$

Thus, nonlinear CD is an input-helicity contrast, while TH DCP is the direct measure of output helicity. Nonlinear chirality transfer bridges these two key measures by suppressing TH generation under pump with resonant-mismatching CP $I^{(3\omega)}_{\text{R}\rightarrow\text{tot}} \ll I^{(3\omega)}_{\text{L}\rightarrow\text{tot}}$, and simultaneously boosting the symmetry-selected RCP component of the TH signal $I^{(3\omega)}_{\boldsymbol{s}^{\text{in}}\rightarrow\text{R}} \gg I^{(3\omega)}_{\boldsymbol{s}^{\text{in}}\rightarrow\text{L}}$. In the following, we focus on the TH signal emitted into the main diffraction channel in transmission geometry under normal-incidence excitation from the air side.

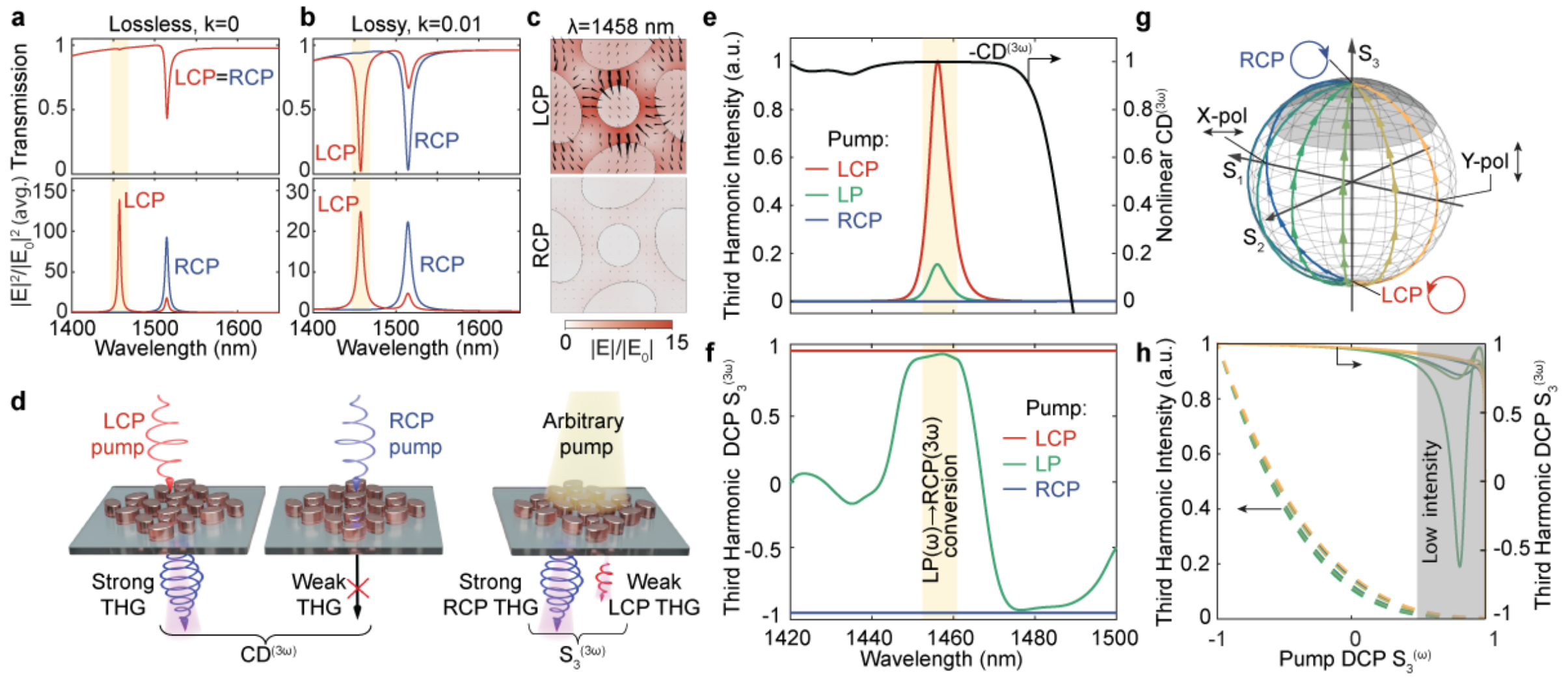

**Figure 3 | Nonlinear chirality transfer and numerical modelling of the chiral metasurface.** **a,b,** Linear transmission and near-field enhancement spectra for lossless and lossy chiral metasurfaces. In both cases, the field enhancement is uniquely determined by the resonance chirality. **c,** Electric-field distributions at the resonant wavelength under resonant (LCP) and nonresonant (RCP) excitation. **d,** Schematic definition of the two key nonlinear figures of merit: nonlinear circular dichroism and the degree of circular polarization of the third-harmonic signal. **e,** Third-harmonic intensity and nonlinear CD spectra under CP and LP excitation. **f,** DCP of the generated TH signal under LCP, linear (LP), and RCP excitation. The spectral region where the TH intensity exceeds 25% of its maximum is highlighted by the yellow band. **g,** Poincaré sphere showing representative trajectories of the pump polarization from LCP to RCP. The shaded region indicates pump states that produce negligible TH intensity. **h,** TH intensity and DCP as functions of pump polarization along the trajectories shown in **g**.

To model the TH response of the metasurface (see Methods for details), we describe the nonlinear polarization of amorphous silicon using an isotropic third-order susceptibility, $\mathbf{P}^{3\omega} = \chi^{(3)}\mathbf{E}(\mathbf{E}\cdot\mathbf{E})$. Figure 3e shows the calculated TH intensity spectra under LCP, LP, and RCP excitation. A sharp resonance appears for both LCP and LP excitation. Notably, the peak TH intensity under LCP pumping is approximately eight times larger than that under LP excitation, since only half of the LP wave power couples to the chiral resonance. In contrast, RCP excitation produces a negligible TH response, leading to near-unity nonlinear CD over a broad wavelength range.

We next examine the polarization of the generated TH signal (Fig. 3f). Under both RCP and LCP excitation, the emitted TH exhibits inverted circular polarization, independent of excitation wavelength, in accordance with symmetry selection rules. In contrast, the TH DCP under LP excitation shows a more intricate behaviour: it reaches near-unity values in the vicinity of the resonance, while outside the resonant region (highlighted by the yellow band in Fig. 3e,f), the TH signal is vanishingly weak. This behaviour indicates that an LP pump generates resonance-enhanced RCP TH emission.

To comprehensively characterize the TH output, we select seven representative trajectories on the Poincaré sphere that span the full space of pump polarizations (Fig. 3g). The corresponding simulated TH intensity and DCP profiles along these paths are shown in Fig. 3h. The results reveal a striking robustness of the TH polarization for all pump states that generate a non-negligible TH response.

Within the effective metamaterial susceptibility framework, the nonlinear response to a pump field with arbitrary polarization, expressed in the circular basis as $\boldsymbol{s}^{\text{in}} = \mathbf{e}_{\text{R}} s_{\text{R}} + \mathbf{e}_{\text{L}} s_{\text{L}} e^{i\Delta\phi}$, can be fully described by four nonzero components of the effective third-order susceptibility tensor $\chi^{(3,\text{eff})}$. The intensities of the LCP and RCP TH components are given by

$$\begin{aligned} I_{\text{L}}^{(3\omega)} &\propto \left|\chi_{\text{LRRR}}^{(3,\text{eff})} s_{\text{R}}^3 + \chi_{\text{LRLL}}^{(3,\text{eff})} s_{\text{R}} s_{\text{L}}^2 e^{2i\Delta\phi}\right|^2, \\ I_{\text{R}}^{(3\omega)} &\propto \left|\chi_{\text{RRRL}}^{(3,\text{eff})} s_{\text{R}}^2 s_{\text{L}} + \chi_{\text{RLLL}}^{(3,\text{eff})} s_{\text{L}}^3 e^{2i\Delta\phi}\right|^2. \end{aligned} \tag{6}$$

A detailed comparison between full-wave simulations and the effective susceptibility description is provided in the Supplementary Section S1. Here, we highlight the central result: at the chiral resonance, the relative magnitudes of the tensor components are $\left|\chi_{\text{LRRR}}^{(3,\text{eff})} : \chi_{\text{LRLL}}^{(3,\text{eff})} : \chi_{\text{RRRL}}^{(3,\text{eff})} : \chi_{\text{RLLL}}^{(3,\text{eff})}\right| = 1:35:28:268$, while all other components of $\chi^{(3,\text{eff})}$ vanish due to the rotational symmetry. Because the resonance selectively couples to and amplifies only the LCP component of the pump, the TH intensity is primarily governed by the pump DCP. The weak intensity variations observed along different trajectories on the Poincaré sphere arise from interference with nonresonant, RCP-induced fields. Dominated by the $\chi_{\text{RLLL}}^{(3,\text{eff})}$ component, the generated TH signal remains nearly perfectly RCP ( $S_3^{(3\omega)} > 0.83$ ) for all pump polarizations spanning the lower three-quarters of the Poincaré sphere.

This behavior constitutes the all-to-circular upconversion effect predicted in our previous work[49], here realized in a **flat**, **single-layer CMOS-compatible** metasurface, representing a highly fabrication-feasible platform.

## Experimental observation of the nonlinear chirality transfer

The chiral metasurface was fabricated using electron-beam lithography followed by inductively coupled plasma reactive-ion etching (see Methods for details). A scanning electron microscope image of the fabricated sample is shown in Fig. 1b. To visualise the modal dispersion, we measured the linear transmission spectra under CP excitation, which are shown in Fig. 4a. All measurement details are provided in the Methods section. One can identify two modes that become degenerate at the Γ-point at 1435 nm. Due to their chirality, the modes only appear as a resonant dip under LCP excitation, while almost vanishing from the RCP transmission. These spectra agree well with the numerically simulated results, shown in Supplementary Fig. S4. Transmission spectra at normal incidence under LCP, LP, and RCP excitation (Fig. 4b) measured using a home-built low-NA transmission microspectroscopy system reveal a pronounced chiral resonance with a Q-factor of 127.

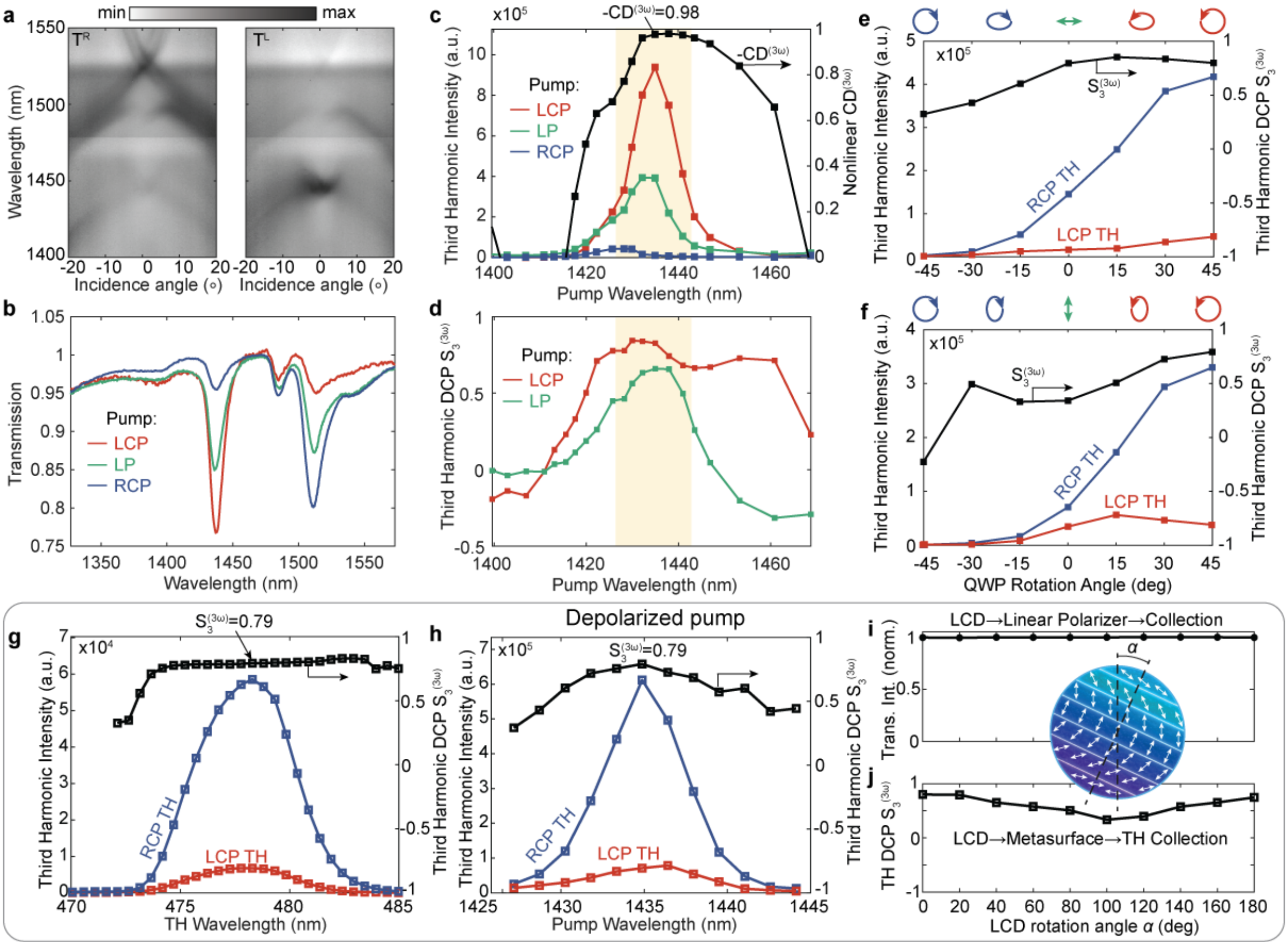

**Figure 4 | Experimental demonstration of nonlinear chirality transfer. a,** Angle-resolved linear transmission spectra under RCP and LCP pump. **b,** Linear transmission spectra at normal incidence under LCP, LP, and RCP excitation. **c,** Measured TH intensity and corresponding nonlinear CD as functions of the pump wavelength. **d,** Measured DCP of the TH signal under LCP and LP excitation. The wavelength range where the TH intensity exceeds 25% of its maximum is highlighted by the yellow band. **e,f,** Intensities of the LCP (red) and RCP (blue) components of the TH signal, together with the TH degree of circular polarization (black). Pump polarization follows Poincaré sphere meridians passing through *x*- and *y*-polarizations. **g,** Intensity of RCP and LCP components of the TH signal, together with the TH DCP under 1435 nm spatially depolarized excitation. **h,** Intensity of RCP and LCP TH components, together with the TH DCP under depolarized excitation as a function of pump wavelength. **i,** Total power of the depolarized beam transmitted through a linear polarizer as a function of the rotation angle of the depolarizer. **j**, TH signal DCP under depolarized excitation as a function of the rotation angle of the depolarizer.

To further validate the TH generation performance of the Si metasurface, a polarization-resolved measurement system was constructed using a near-infrared wavelength-tunable femtosecond laser. The experimentally measured TH signals (Fig. 4c) show good agreement with the simulations in Fig. 3e. Under LCP and LP excitation, the TH intensity as a function of the pump wavelength shows a resonant peak with a full width at half-maximum of $\Delta\lambda(I^{(3\omega)}) = 10$ nm, which corresponds to the effective nonlinear Q-factor $Q^{NL,eff} = \frac{\lambda}{\Delta\lambda(I^{(3\omega)})} \approx 140$, where $\lambda$ is the pump wavelength. At the same time, the resonant-to-nonresonant intensity ratios (at $\lambda = 1435$ nm and $\lambda = 1407$ nm) exceed 100 and 35 for LCP and LP pump, respectively.

In contrast, RCP excitation produces a negligible TH signal. This pronounced asymmetry gives rise to a near-unity nonlinear circular dichroism of $\mathrm{CD}^{(3\omega)} = -0.98$. The measured DCP of the nonlinear signal (Fig. 4d) confirms the helicity-inversion effect under LCP pumping with good quantitative agreement: the TH DCP reaches $S_3^{(3\omega)} = 0.83$ at resonance. The experimentally observed deviation of $S_3^{(3\omega)}$ from unity can be attributed to fabrication-induced symmetry breaking, as well as to deviations of the excitation beam from an ideal circularly polarized plane wave. Under LP excitation, the DCP clearly approaches that obtained under LCP pumping, reaching $S_3^{(3\omega)} = 0.66$ at resonance. Thus, the generation of a highly CP TH signal is consistently observed across the entire resonant band.

Next, we experimentally characterize the nonlinear response along two representative meridians on the pump-polarization Poincaré sphere. To probe the meridians of the sphere, the pump light is passed through a linear polarizer followed by a quarter-wave plate (QWP). Along the first meridian, where the pump polarization evolves from RCP through $x$-polarized light to LCP, both RCP and LCP components of the TH signal increase monotonically, with the RCP component consistently dominating (Fig. 4e). Notably, an exceptionally high TH DCP of $S_3^{(3\omega)} = 0.8$ is achieved under $x$-polarized excitation, exceeding the value reported in Fig. 4d. We attribute this enhancement to the fact that the LP pump used in Fig. 4d was oriented at a 15° angle relative to the $x$-axis. Along the second meridian, corresponding to a polarization evolution from RCP through $y$-polarized light to LCP, the TH intensity and DCP exhibit a slightly different trend (Fig. 4f), which may be associated with accidental symmetry breaking due to the fabricated nanopattern of the sample or excitation beam asymmetry.

To further verify that the observed response arises from conventional perturbative third-order nonlinearity, we perform power-dependent measurements using a narrowband pump pulse centered at 1435 nm, within the resonant band (see Supplementary Fig. S8). Under LCP excitation, the TH intensity follows the expected power-law dependence on the pump intensity, $I^{(3\omega)} \propto (I^{(\omega)})^n$. The extracted exponent $n = 2.85$ is in reasonable agreement with the theoretical value of 3. Although anomalous power-law behaviour has been recently reported in nonlinear metasurfaces[63], no pronounced deviations are observed in our system.

Finally, we demonstrate that even a spatially depolarized coherent beam can be used to excite the resonance and produce a nonlinear chiral signal. A linearly polarized laser beam passes through a liquid crystal depolarizer Thorlabs DPP25-C, that imprints a rapidly varying spatial distribution of polarization onto the cross-section of the beam, which is later focused on the metasurface using a lens (see schematics in Fig. 1d). The polarization of the output beam is rotated by 4° per 25 μm-wide strip, thus covering all possible linear polarizations multiple times across the 5 mm-wide beam cross-section. While the electric field of this beam retains its spatial coherence, the rapid spatial variation of the polarization allows one to use it as an effectively depolarized excitation source.

When the light pulse is centered at 1435 nm, the output nonlinear signal is dominated by its RCP component, featuring TH DCP of $S_3^{(3\omega)} = 0.79$ (Fig. 4g). This DCP is consistently observed across the whole nonlinear response linewidth. As we scan the pump wavelength in the vicinity of the resonance, the integrated TH signal remains strongly RCP, exhibiting $S_3^{(3\omega)} \in [0.57, 0.79]$ at all wavelengths for which the TH intensity exceeds 25% of the maximum (Fig. 4h), confirming robustness of the proposed symmetry-based mechanism for the nonlinear chirality transfer.

The absence of a net linear-polarization bias in the spatially integrated beam can be easily verified by measuring the beam power after it passes through a linear polarizer. As we rotate the depolarizer, the transmitted intensity exhibits variations of less than 1% (Fig. 4i). However, rotation of the depolarizer may noticeably affect the TH DCP of the TH signal, generated by the metasurface excited at the chiral resonance (Fig. 4j). While the theoretical model of TH generation under depolarized beam excitation, described in Supplementary Sections 4 and 5, only predicts reduction of $S_3^{(3\omega)}$ with increased aperture of the TH collection objective with no dependence on depolarizer rotation, we notice that nonlinear processes are highly sensitive to even small variations of the local fields. Slight asymmetry of the fabricated metasurface lattice, asymmetry of the excitation beam, and opposite transverse phase gradients introduced by the depolarizer (Supplementary Section 5) can lead to significant corrections to the polarization of the generated TH signal. Nevertheless, despite the complex spatially varying polarization excitation, the TH DCP remains above 0.33 for all rotation angles and even reaches 0.8, unambiguously confirming the dominance of the RCP TH component.

Benchmarked against the existing literature in Fig. 1c, the simultaneous realization of high TH DCP, near-unity nonlinear CD, and a high effective Q-factor demonstrates the effectiveness of complementary symmetry breaking and symmetry preservation as a design principle for controlling nonlinear optical responses, particularly for achieving strong chirality.

## Conclusion

We have experimentally demonstrated a symmetry-based mechanism that enables chirality selection and polarization purification in nonlinear all-dielectric metasurfaces. The central element of this mechanism is a degenerate pair of high-Q resonances with maximal chirality, which selectively amplifies one circular component of the pump and, enforced by rotational symmetry, converts it into a nonlinear signal of opposite handedness. This process realizes nonlinear chirality transfer, whereby a chiral resonant response at the pump frequency deterministically produces a circularly polarized harmonic output. Although our experimental demonstration focuses on third-harmonic generation, the underlying principle is universal: our earlier theoretical work[49] established its applicability to both second- and third-harmonic processes, with natural extensions to higher-order nonlinear interactions.

From an experimental perspective, our metasurface integrates several advances that have not been achieved simultaneously in prior studies. In the linear regime, we demonstrate a maximally chiral resonance in a rotationally symmetric dielectric structure enabled by the mechanism of quasi-guided modes. Previous experimental realizations of chiral quasi-BICs relied on architectures lacking rotational symmetry and were therefore incompatible with nonlinear chirality transfer. In contrast, our approach provides a practical route to realizing chiral high-Q modes in a single-layer, flat, CMOS-compatible platform, without multilayer stacking, multi-height patterning, or substrate engineering.

In the nonlinear regime, we report a near-unity nonlinear circular dichroism of −0.98 together with a degree of circular polarization of the harmonic signal reaching 0.83 under circularly polarized excitation and 0.8 under linearly polarized excitation. These values are among the highest reported for dielectric nonlinear metasurfaces and have not previously been achieved jointly within a single high-Q platform. Importantly, nonlinear circular dichroism alone, as commonly reported for conventional chiral structures, merely reflects intensity asymmetry and provides no information about the polarization state of the generated light. By simultaneously achieving near-unit nonlinear circular dichroism and a high degree of circular polarization, we demonstrate deterministic control over the functional polarization state of the nonlinear output, a prerequisite for applications requiring a deterministic output helicity. Notably, the nonlinear harmonic preserves its handedness even under spatially depolarized coherent light excitation.

Our simulations further reveal that in the ideal case, pump states spanning approximately three-quarters of the Poincaré sphere generate a stable, right-handed third-harmonic output, while the remaining quadrant yields a strongly suppressed response. This behavior is consistent with symmetry selection rules for rotationally symmetric nonlinear processes, but it goes substantially beyond them: the presence of chirality stabilizes the nonlinear polarization output across a broad range of input polarization states. This robustness highlights the potential of chiral high-Q modes as nonlinear chiral emitters, capable of enforcing the handedness of the generated harmonics independently of the pump polarization.

In contrast to a recent study[48] demonstrating control over the degree of circular polarization of the nonlinear signal in achiral metasurfaces, our mechanism relies on intrinsic structural symmetry rather than fine-tuned modal intersections, suggesting broad applicability and robustness. Introduction of an all-optically controlled degree of freedom, for example, strongly asymmetric excitation[64] or intensity-induced modulation[65], may be used to modify the nonlinear selection rules, enabling nonlinear logic functionalities in ultrathin nonlinear photonic metadevices.

In summary, our work demonstrates how symmetry of resonant metasurfaces can be systematically leveraged to control not only the intensity but also the functional polarization of nonlinear optical signals. By combining chirality with rotational symmetry, we establish a general design paradigm for symmetry-controlled nonlinear chiral devices. This approach opens new avenues toward compact sources of structured nonlinear light, ultrathin polarization logic, spin–orbit–engineered nonlinear photonics and quantum nonlinear metasurfaces with chirality-controlled entanglement.

## Methods

**Scattering matrix bandstructure calculations:** The simulations were performed using the rigorous coupled-wave analysis (RCWA) open-source package Inkstone[66]. The unit cell was defined using the two lattice vectors and cylindrical pillars with circular base faces. We defined the refractive index of the silicon as 3 and the silica substrate refractive index as 1.45, the upper half-space was air with a refractive index 1. The metasurface parameters are period $p = 955$ nm, thickness $H = 320$ nm, larger pillar radius $R = 240$ nm, and a smaller pillar radius $R_c = 150$ nm. The scattering matrix for the open radiative channels was calculated for various Bloch vector values and frequencies; its norm on a logarithmic scale is presented in Fig. 2b,e to visualise the band structure below and above the light cone. RCWA, being a computational algorithm that relies on the Fourier transformation of the material properties and fields, is a fast and accurate tool for modelling periodic metasurfaces with simple shapes, especially in the vicinity of Rayleigh anomalies (openings of diffraction channels).

**COMSOL eigenmode simulations:** Resonant modes of the metasurface were computed using the COMSOL eigenmode solver. Periodic boundary conditions along the lateral $x$- and $y$-directions and ports on the upper and lower boundaries were used to define the simulation domain. The metasurface parameters were the same as for the bandstructure calculations. The chirality of a mode with electric field $\mathbf{E}(\mathbf{r})$ was evaluated based on the far-field polarization of the mode on the air side:

$$\mathbf{E}_{\text{rad}} = \frac{1}{S_{\text{u.b.}}} \int_{S_{\text{u.b.}}} \mathbf{E}(\mathbf{r})\, dS, \tag{7}$$

where $S_{\text{u.b.}}$ is the upper boundary of the metasurface. Accordingly, the DCP, or the normalized third Stokes parameter of the radiation, used as a measure of a mode's chirality, was evaluated as $S_3 = 2\,\text{Im}(\text{E}_{\text{rad},x}\text{E}^*_{\text{rad},y})/\left(\left|\text{E}_{\text{rad},x}\right|^2 + \left|\text{E}_{\text{rad},y}\right|^2\right)$.

**Temporal coupled-mode theory for chiral modes of E representation of $C_4$ group:** Consider a degenerate pair of modes at the Γ-point, with electric fields satisfying Eq. (1). Within the temporal coupled-mode theory[67], the interaction of such a pair of modes with the radiation continuum at normal incidence is described by equation

$$\frac{d\boldsymbol{a}}{dt} = [i\omega_0 - \gamma]\boldsymbol{a} + \widehat{\boldsymbol{K}}\boldsymbol{s}^{\text{in}}, \tag{8}$$

where $\boldsymbol{a} = (a_1, a_2)^{\text{T}}$ is the vector of modal amplitudes, $\omega_0 + i\gamma$ is the complex resonant frequency of the modes, $\widehat{\boldsymbol{K}}$ is the matrix of resonant coupling constants that defines coupling to the amplitudes of incoming waves $\boldsymbol{s}^{\text{in}} = (s^{\text{in}}_{1,x}, s^{\text{in}}_{1,y}, s^{\text{in}}_{2,x}, s^{\text{in}}_{2,y})^T$, with subscripts 1,2 denoting the upper and lower half-spaces and subscripts $x$ and $y$ denoting the Cartesian components of wave polarization. Due to $C_4$ symmetry and transformation properties of the modes Eq. (1), the coupling constants of the two modes are related through 90-degree rotation operation, and take the form

$$\widehat{\boldsymbol{K}} = \begin{pmatrix} \boldsymbol{K}_1 \\ \boldsymbol{K}_2 \end{pmatrix} = \begin{pmatrix} K_{1,x} & K_{1,y} & K_{2,x} & K_{2,y} \\ -K_{1,y} & K_{1,x} & -K_{2,y} & K_{2,x} \end{pmatrix}, \tag{9}$$

where $\boldsymbol{K}_{1,2}$ are the vectors of resonant coupling. Now, assuming mode 1 is tuned to selectively interact with LCP light incident from the top, i.e., it is fully decoupled from RCP excitation $\boldsymbol{s}^{\text{in}} \propto (1, i, 0,0)^{\text{T}}$, we obtain $K_{1,y} = iK_{1,x}$. Under such circumstances, any excitation from the top $\boldsymbol{s}^{\text{in}} = (s^{\text{in}}_{1,x}, s^{\text{in}}_{1,y}, 0,0)^{\text{T}}$ at frequency $\omega$ results in the stationary solution $\boldsymbol{a}(t) = \boldsymbol{a}e^{i\omega t}$

$$\boldsymbol{a} = \frac{\widehat{\boldsymbol{K}}\boldsymbol{s}^{\text{in}}}{[i(\omega - \omega_0) + \gamma]} = \frac{\sqrt{2}K_{1,x}s^{\text{in}}_{1,-}}{[i(\omega - \omega_0) + \gamma]}\begin{pmatrix} 1 \\ -i \end{pmatrix}, \tag{10}$$

where $s^{\text{in}}_{1,-} = {}^{1}/_{\sqrt{2}}(s^{\text{in}}_{1,x} + is^{\text{in}}_{1,y})$ is the amplitude of the LCP component of the incident wave coming from top. We conclude that regardless of the incident light polarization, both modes are excited with the same absolute value of the amplitude and a constant phase offset $a_1 = ia_2$; pure RCP excitation does not resonantly excite the doublet. Hence, omitting weak nonresonant terms, the induced field within the metasurface $\mathbf{E}^{\omega}(\mathbf{r}) \approx a_1\mathbf{E}_1(\mathbf{r}) + a_2\mathbf{E}_2(\mathbf{r})$ obeys the chiral transformation rule Eq. (2).

**Semi-analytical perturbation model:** We consider the electric-field distribution $\mathbf{E}(\mathbf{r}) = \mathbf{E_1}(\mathbf{r})$ of one of the degenerate resonant modes of the unperturbed structure. The other mode is related to the first one through Eq. (1). When a boundary-shift perturbation is introduced, the perturbed resonant field $\tilde{\mathbf{E}}(\mathbf{r})$ remains unchanged in regions where the permittivity is preserved,

$$\tilde{\mathbf{E}}(\mathbf{r}) = \mathbf{E}(\mathbf{r}) \text{ for } \mathbf{r} \in U_{\mathrm{res}} \cup U_{\mathrm{bg}}.$$

In regions where the permittivity is modified, we employ the following approximation:

$$\tilde{\mathbf{E}}_{||}(\mathbf{r}) = \mathbf{E}_{||}(\mathbf{r})\Big|_{\mathbf{r}\in\delta U_{\mathrm{res}} \cup \delta U_{\mathrm{bg}}},$$
$$\tilde{\mathbf{E}}_{\perp}(\mathbf{r}) = \frac{\mathbf{E}_{\perp}(\mathbf{r})}{\varepsilon_{\mathrm{Si}}}\Bigg|_{\mathbf{r}\in\delta U_{\mathrm{res}}} \text{ and } \tilde{\mathbf{E}}_{\perp}(\mathbf{r}) = \varepsilon_{\mathrm{Si}}\mathbf{E}_{\perp}(\mathbf{r})\Big|_{\mathbf{r}\in\delta U_{\mathrm{bg}}}, \quad (11)$$

where $\perp$and $||$ denote the normal and tangential components with respect to the local surface of an elliptical cylinder passing through the point $\mathbf{r}$. These surfaces form a continuous morphing family connecting the circular and elliptical pillars, parameterized by $\Delta R' \in [0, \Delta R]$ and defined by the rotation angle $\theta$ and half-lengths $R \pm \Delta R'$. Unlike the unperturbed field $\mathbf{E}(\mathbf{r})$, the perturbed field $\tilde{\mathbf{E}}(\mathbf{r})$ is constructed to satisfy the relevant electric-field interface continuity conditions for the modified structure.

Using the reciprocity principle[68], we evaluate the coupling coefficients $K_{1,\alpha}$ $(\alpha = x, y)$ of the perturbed resonant mode as

$$K_{1,\alpha} \propto \int \left(\tilde{\varepsilon}(\mathbf{r}) - \varepsilon_{\mathrm{bg}}(\mathbf{r})\right) \mathbf{E}_{\alpha}^{\mathrm{bg}}(\mathbf{r})\tilde{\mathbf{E}}(\mathbf{r})d^3\mathbf{r}, \quad (12)$$

where $\varepsilon_{\mathrm{bg}}(\mathbf{r})$ is the background permittivity in the absence of the resonator, and $\mathbf{E}_{\alpha}^{\mathrm{bg}}(\mathbf{r})$ is the background electric field generated by a plane wave with polarization $\alpha$ incident in the reciprocal configuration. In essence, the original radiation problem is recast as a reciprocal scattering problem.

To account for the substrate with permittivity $\varepsilon_{\mathrm{SiO_2}} = n_{\mathrm{SiO_2}}^2 = 1.45^2$, we set $\varepsilon_{\mathrm{bg}}(z \le 0) = \varepsilon_{\mathrm{SiO_2}}$and $\varepsilon_{\mathrm{bg}}(z > 0) = 1$. The corresponding background field in the reciprocal problem is

$$\mathbf{E}_{\alpha}^{\mathrm{bg}}(\mathbf{r}) = \hat{\mathbf{e}}_{\alpha}\left(\exp(ik_0 z) + \tfrac{1-n_{\mathrm{SiO_2}}}{1+n_{\mathrm{SiO_2}}}\exp(-ik_0 z)\right),$$

where $\hat{\mathbf{e}}_{\alpha}$ is the polarization unit vector.

Substituting the perturbed field expansion from Eq. (11) into Eq. (12), we obtain

$$K_{1,\alpha} \propto \int_{U_{\mathrm{res}}} \mathbf{E}_{\alpha}^{\mathrm{bg}}(\mathbf{r})\mathbf{E}(\mathbf{r})d^3\boldsymbol{r} + \int_{\delta U_{\mathrm{res}}} \mathbf{E}_{\alpha}^{\mathrm{bg}}(\mathbf{r})\left(\mathbf{E}_{||}(\mathbf{r}) + \frac{\varepsilon(\mathbf{r})}{\tilde{\varepsilon}(\mathbf{r})}\mathbf{E}_{\perp}(\mathbf{r})\right)d^3\boldsymbol{r}. \quad (13)$$

While the resonant fields $\mathbf{E}(\mathbf{r})$ and $\tilde{\mathbf{E}}(\mathbf{r})$ can be properly normalized to yield exact equalities[60,69], this is of minor importance for the evaluation of the far-field polarization state of the resonant mode. Finally, chirality of the mode in this interface-perturbative approach is evaluated as $S_3 = 2\,\mathrm{Im}(K_{1,x}K_{1,y}^*)/\left(|K_{1,x}|^2 + |K_{1,y}|^2\right)$.

**Numerical modeling: Linear vs. nonlinear chiral response:** Plane wave scattering in the linear regime was modeled in the "Electromagnetic Waves, Frequency Domain" module of COMSOL. The boundary conditions and unit cell design were adopted from the eigenmode simulation, with a small imaginary part of the Si refractive index of 0.01 introduced to mimic the material losses. To ensure simulation accuracy in the following nonlinear simulation, the spatial mesh in the Si, $SiO_2$, and air domains was defined explicitly with the largest element size not exceeding $^{0.17\lambda}/_{3n}$, where $\lambda = 1400$ nm is the minimum pump wavelength and $n$ is the refractive index of a particular material. Transmission coefficients were extracted using the port functionality.

The nonlinear part of the simulation was done in the second consecutive simulation, where the calculated electric fields in the linear scattering simulation were used as a source of nonlinear polarization. The material properties were kept the same in both the linear and idealized nonlinear simulations. The ports at the upper and lower boundaries were replaced with second-order scattering conditions; additionally, the frequency of the simulation was explicitly defined to be 3 times higher than the frequency of the first simulation. No incoming waves were used in this simulation. The far-field TH radiation in the main diffraction channel was extracted at the lower boundary of the simulation domain according to Eq. (7).

**Fabrication of the chiral metasurface:** First, a 320-nm-thick α-Si film was deposited on a $SiO_2$ substrate. Hydrogen silsesquioxane (HSQ, Dow Corning XR-1541-006) resist was spin-coated at 5000 rpm to obtain a ∼150 nm layer, followed by spin coating of an ∼80 nm AR-PC 5090 conductive layer at 2000 rpm. The sample was patterned by electron-beam lithography (EBL, Raith Eline Plus, 30 keV, 450 pA). The conductive layer was rinsed off with deionized water for 1 min, and the HSQ pattern was developed in 25 wt.% TMAH for 3 min, followed by a deionized water rinse and isopropanol drying. The α-Si layer was etched using ICP-RIE (Oxford Plasmalab System 100) with $Cl_2$ gas (22 sccm), ICP power 300 W, RF power 100 W, chamber pressure 5 mTorr, and chiller temperature 10 °C.

**Linear optical characterization:** The linear transmission of the metasurface was measured in a home-built setup (Supplementary Fig. S9) using a broadband incoherent light source (Thorlabs OSL2, 1300–1600 nm). The incident beam was converted to circular polarization using a linear polarizer and a quarter-wave plate. To address the angle sensitivity of the quasi-guided modes, a long-focal-length NIR lens (f = 50 mm) generated a low-NA excitation beam near normal incidence. Transmitted light was collected with a 10× objective (NA = 0.26) and analyzed using a spectrometer (Kymera 328i) coupled to a cooled InGaAs CCD. Angle-resolved transmission spectra were measured using a commercial microscope system, where the transmitted light was dispersed by a diffraction grating and recorded by a NIR CMOS camera.

**Nonlinear optical characterization (THG):** For third-harmonic generation (THG) measurements, the incoherent light source was replaced with a home-built wavelength-tunable optical parametric amplifier (OPA) used in Ref. [70] pumped by a pulsed laser (Amplitude Yuja, center wavelength of 1030 nm, repetition rate of 100 kHz, and pulse duration of 400 fs). The output wavelength of the OPA was tuned from 1400 to 1470 nm by selecting periodically poled lithium niobate (PPLN) crystals with different poling periods and adjusting the crystal temperature. The pump polarization was controlled using a linear polarizer with a quarter-wave plate to tune the $S_3$ of the excitation beam, or a half-wave plate for achieving arbitrary linear polarization. THG emission was collected using the same 10× objective and analyzed with a spectrometer (Kymera 328i) coupled to a Si CCD. A short-pass filter blocked the pump light, and RCP/LCP components of the THG signal were resolved using a quarter-wave plate followed by a linear polarizer (Supplementary Fig. S10).

**AI and LLM Usage:** During the preparation of this manuscript, the authors used OpenAI's ChatGPT-5 and ChatGPT-6 for formatting, text editing, and proofreading to improve readability and grammatical accuracy.

## Authors' Contributions

D.G., C.W.Q., and L.W. conceived the project idea. Q.W., J.K.W.Y., C.W.Q., and L.W. supervised the project. D.G., S.T., C.W.Q., and L.W. developed the theoretical background. D.G. performed numerical simulations. C.F.P., H.T.W., B.W., S.Z., and H.W. fabricated the samples. C.F.P. prepared the SEM images. W.J. and X.M. designed the optical experiments and built the linear and TH characterization setups. W.J., X.M., W.C., Z.X., and C.F.P. performed the experiments. W.J., D.G., X.M., W.C., C.F.P., and Z.X. analysed the data. D.G., W.J., C.W.Q., and L.W. wrote the manuscript with the contribution of all authors.

## Competing interests

The authors declare no competing interests.

## Data availability

All data supporting the findings of this study are available within the Article and its Supplementary Information. Additional data are available from the corresponding authors upon reasonable request.

## Code availability

No custom code is required to reproduce the findings of this study. The numerical methods and software used in the study are described in the Methods section.

## Acknowledgements

We thank Dr. Jianfeng Chen (NUS), Dr. Jieyuan Cui (NTU), and Ms. Hanyu Liu (NTU) for helpful discussions. This work was supported by the National Research Foundation, Singapore (NRF-CRP26-2021-0004, NRF-CRP31-0007, NRF2023-ITC004-001, NRF-CRP29-2022-0003, and NRF-MSG-2023-0002), the Ministry of Education, Singapore (MOE-T2EP50223-0001, MOE-MOET32024-0005, MOET32024-0001), the Agency for Science, Technology and Research (A*STAR) (MTC IRG M24N7c0083, Programmatic Funds R23I0IR041, M23M2b0056), the Singapore University of Technology and Design (SUTD) under the Kickstarter Initiative (SKI 2021-04-12), the Advanced Research and Technology Innovation Centre (ARTIC), the National University of Singapore under Grant (project number: AFP-RP4). D.G. and C.P. acknowledge the support of the Ph.D. Research Scholarship Scheme.